\documentclass[reprint,amsmath,amssymb,aps]{revtex4-1}
\usepackage{graphicx}
\usepackage{epstopdf}

\begin{document}
\title{Measurement-induced operation of two-ion quantum heat machines}
\author{Suman Chand}
 \email{suman.chand@iitrpr.ac.in}
\author{Asoka Biswas}%
\affiliation{%
Department of Physics, Indian Institute of Technology Ropar, Rupnagar, Punjab 140001, India
}%
\date{\today}%

\begin{abstract}
We show how one can implement a quantum heat machine by using two interacting trapped ions, in presence of a thermal bath. The electronic states of the ions act like a working substance, while the vibrational mode is modelled as the cold bath. The heat exchange with the cold bath is mimicked by the projective measurement of the electronic states. We show how such measurement in a suitable basis can lead to either a quantum heat engine or a refrigerator, that undergoes a quantum Otto cycle. The local magnetic field is adiabatically changed during the heat cycle. The performance of the heat machine depends upon the interaction strength between the ions, the magnetic fields, and the measurement cost. In our model, the coupling to the hot and the cold baths are never switched off in an alternative fashion during the heat cycle, unlike other existing proposals of quantum heat engines. This makes our proposal experimentally realizable using current tapped-ion technology.
\end{abstract}

\maketitle

\section{\label{sec:i}Introduction\protect\\   }
In recent years, the study of quantum thermodynamics \cite{Quan-Thermodynamics book G. Mahler,Quan-Thermodynamics book G. P. Beretta} has attracted a lot of attention to understand the fundamental relation between quantum mechanics and thermodynamics \cite{Quan-Thermodynamics cycle}. In this context, the concept of quantum heat engines (QHEs) was first introduced by Scovil and Schulz-Dubois using three-level masers \cite{Scovil maser paper}.  Since then, a significant amount of effort has been devoted in studies of several quantum heat machines, including different heat engines \cite{Scully Science,kosloff,Quantum Otto engine of a two-level atom with single-mode fields,Entangled QHE based on two two spin systems with DM,Four-level entangled quantum heat engines,Thermal entanglement in two-atom cavity QED and the entangled quantum Otto engine,johal,Effects of reservoir squeezing on quantum systems and work extraction,Atlintas,Thermal entangled four level QOE,A Special entangled QHE based on two qubit Heisenberg XX model,Special coupled quantum otto cycle,Thermal entangled QHE working with 3-qubit XX model,Lipkin-Meshkov-Glick Model in QOE,Performence of coupled system as QHE,QHE with multilevel quantum system-Quan,Bender multi level system,Quantum harmonic oscillator Kosloff,Efficiency at maximum power of a quantum heat engine based on two coupled oscillators,Quantum-classical transition of photon-Carnot engine induced by quantum decoherence-Quan,single ion-Abah,optomechanical system-Meystre,Magnon-driven quantum-dot heat engine,Isolated quantum heat engine- bosonic system,Mustecaphoglu Sci Rep,Quantum Brayton cycle with coupled systems as working substance,QHE: Multiple-State 1D Box system,Stirling engine}, namely, Carnot \cite{Scully Science,Bender multi level system,Quantum-classical transition of photon-Carnot engine induced by quantum decoherence-Quan}, Otto \cite{Effects of reservoir squeezing on quantum systems and work extraction,Quantum Otto engine of a two-level atom with single-mode fields,Thermal entanglement in two-atom cavity QED and the entangled quantum Otto engine,johal,Atlintas,Thermal entangled four level QOE,A Special entangled QHE based on two qubit Heisenberg XX model,Special coupled quantum otto cycle,Lipkin-Meshkov-Glick Model in QOE,Performence of coupled system as QHE,Quantum harmonic oscillator Kosloff,Efficiency at maximum power of a quantum heat engine based on two coupled oscillators,single ion-Abah,Mustecaphoglu Sci Rep}, Brayton \cite{Quantum Brayton cycle with coupled systems as working substance,QHE: Multiple-State 1D Box system}, Diesel \cite{QHE: Multiple-State 1D Box system} and Stirling \cite{Stirling engine} and 
also  the refrigerator \cite{Refrigerator spin 1/2 system,Carnot like refrigerator, Otto Refrigerator with squeezing, Abah Otto Refrigerator, Stirling engine, Transitions between refrigerator regions in extremely short cycle, The quantum refrigerator in a two qubit XXZ Hisenberg model,A thermal entangle quantum refrigerator based on a two qubit Hisenberg model with DM interection in a external magnetic field,Entanglement enhances cooling in microscopic quantum refrigerators}.  


QHEs have been proposed using different working substances, e.g., two-level systems \cite{Quan-Thermodynamics cycle,kosloff,Quantum Otto engine of a two-level atom with single-mode fields,Entangled QHE based on two two spin systems with DM,Four-level entangled quantum heat engines,Thermal entanglement in two-atom cavity QED and the entangled quantum Otto engine,johal,Effects of reservoir squeezing on quantum systems and work extraction,Atlintas,Thermal entangled four level QOE,A Special entangled QHE based on two qubit Heisenberg XX model,Special coupled quantum otto cycle,Thermal entangled QHE working with 3-qubit XX model,Lipkin-Meshkov-Glick Model in QOE,Performence of coupled system as QHE}, multi-level systems \cite{QHE with multilevel quantum system-Quan,Bender multi level system}, and harmonic oscillators \cite{Quan-Thermodynamics cycle,Performence of coupled system as QHE,Quantum harmonic oscillator Kosloff,Efficiency at maximum power of a quantum heat engine based on two coupled oscillators}. 
Several proposals have been made to implement such engines in cavity QED \cite{Quantum-classical transition of photon-Carnot engine induced by quantum decoherence-Quan}, single ion \cite{single ion-Abah}, optomechanical systems \cite{optomechanical system-Meystre}, quantum dots \cite{Magnon-driven quantum-dot heat engine}, and cold bosons \cite{Isolated quantum heat engine- bosonic system}. 

In a standard heat engine, a working substance extracts heat from a hot bath at an equilibrium temperature $T_H$, does a certain amount of work $W$, and then releases the rest of the energy to the cold bath at an equilibrium temperature $T_L (<T_H)$. The ideal Carnot engine sets an upper limit for the efficiency of such an engine at $\eta_C=1-T_L/T_H$. Several authors have investigated the performance of the QHEs to determine whether the quantum nature of the associated heat baths provides any advantage over their classical counterparts  \cite{R. Uzdin PRX,Scully Science}. For example, QHEs can operate with an efficiency beyond the classical Carnot bound $\eta_C$ without violating the second law of thermodynamics by using quantum coherent heat reservoirs \cite{Scully Science,Mustecaphoglu Sci Rep} or the squeezed heat bath \cite{Effects of reservoir squeezing on quantum systems and work extraction}.

Entanglement \cite {EPR paper,Schrdinger paper,Bennet Nature} represents a nonclassical nonlocal correlation between two or more quantum systems that does not have any classical counterpart.  It is quite interesting to investigate how the entanglement in the working substance affects the basic quantum thermodynamical quantities, namely work and heat. In fact, an entangled system is more efficient in extracting work than the system without such nonclassical properties \cite{kosloff,Entangled QHE based on two two spin systems with DM,Four-level entangled quantum heat engines,Thermal entanglement in two-atom cavity QED and the entangled quantum Otto engine,johal,Effects of reservoir squeezing on quantum systems and work extraction,Atlintas,Thermal entangled four level QOE,A Special entangled QHE based on two qubit Heisenberg XX model,Special coupled quantum otto cycle,Thermal entangled QHE working with 3-qubit XX model,Lipkin-Meshkov-Glick Model in QOE,Performence of coupled system as QHE,Quantum Brayton cycle with coupled systems as working substance}. In this context, different types of interaction between the subsystems of the working substance have been employed, namely, Heisenberg XXX \cite{Four-level entangled quantum heat engines,johal, Atlintas,Thermal entangled four level QOE} and  XX interaction \cite{Entangled QHE based on two two spin systems with DM,Thermal entanglement in two-atom cavity QED and the entangled quantum Otto engine,A Special entangled QHE based on two qubit Heisenberg XX model,Thermal entangled QHE working with 3-qubit XX model,Quantum Brayton cycle with coupled systems as working substance}, Dzyaloshinski-Moriya interactions \cite{Entangled QHE based on two two spin systems with DM}, and squeezing \cite{Atlintas},  to show that the engine efficiency can be a function of the entanglement prevailing in the system. In this paper, we demonstrate how a quantum heat machine can be implemented using two ions. The thermal environment works as a hot bath, while the common vibrational mode of the ion is made to work like a cold bath. We explore the effect of the coupling between the electronic states of the two ions on the efficiency of the heat engine. We discuss the suitable strategy such that the same system can also perform like a refrigerator.

In a standard classical heat engine, 
the working substance interacts with the hot bath and the cold bath in an alternative fashion. This assumes the ability to selectively switch off or switch on the coupling with the bath during certain strokes of the heat cycle. In all the existing proposals, as mentioned above, primary efforts have been made to directly map such classical heat strokes into quantum heat engines.  However, such a `reciprocating' cycle may not be feasible in quantum regime, as the working substance experiences an always-on interaction with the bath \cite{kurizki,kurizki1}. In our model, we show that it is rather possible to switch between the two baths, as required in heat cycles, in presence of such an always-on interaction. In this context, we propose use of the projective measurement of the electronic states of the ions in suitable basis, that leads to an effective heat exchange with the cold bath. Further, suitable choice of projected states can lead to either a heat engine or a refrigerator cycle. 
In view of the above, the two-ion system, as we describe next, poses as an experimentally feasible model to implement a quantum heat machine.

The paper is organized as follows. In Sec. II, we describe our two-ion model and discuss how the quantum heat machines can be implemented in such system. We conclude the paper in Sec. III.

\section{Implementation of the cycles of quantum heat machines}
\subsection{Model}\label{s:ii}
We consider two trapped two-level ions with the lowest lying electronic states $|\pm\rangle$ as the relevant energy levels.  These internal states of the ions interact with a common vibrational mode $a$. 
The Hamiltonian that describes this system can be written as (in unit of Planck's constant $\ensuremath{\hbar=1}$)
\begin{equation}
H_{1}=H_{S}+H_{{\rm ph}}+H_{{\rm int}}\;,
\end{equation}
where
\begin{eqnarray}
H_{S} & = & J\left(\sigma_{+}^{(1)}\sigma_{-}^{(2)}+\sigma_{-}^{(1)}\sigma_{+}^{(2)}\right)+B\left(\sigma_{z}^{(1)}+\sigma_{z}^{(2)}\right)\;, \nonumber \\
 H_{\rm ph} & = & \omega a^{\dagger}a\;, \nonumber \\
H_{\rm int} & = & k_{1}\left(a^{\dagger}\sigma_{-}^{(1)}+\sigma_{+}^{(1)}a\right)+k_{2}\left(a^{\dagger}\sigma_{-}^{(2)}+\sigma_{+}^{(2)}a\right)\;.
\end{eqnarray}
Here $H_S$ represents the unperturbed Hamiltonian of two ions, which interact with each other with the corresponding coupling constant $J$ (as in the Heisenberg XX model), $H_{\rm ph}$ is the energy of the vibrational mode with frequency $\omega$, and  $H_{\rm int}$ defines the interaction between the internal and the vibrational degrees of freedom of the ion. The interaction strength between the electronic transitions of the $i$th ion and the vibrational mode is given by $k_{i}$, ($i\in 1,2$). A magnetic field of strength $B$ is applied along the quantization axis. The cases $J>0$ and $J<0$ correspond to the antiferromagnetic and the ferromagnetic interactions, respectively. In this paper, we choose the antiferromagnetic case only.

We consider the electronic states of the two-ion joint system as the working substance S of our heat machine. In the joint basis of the two ions, $\{\left|++\right\rangle ,\left|+-\right\rangle ,\left|-+\right\rangle ,\left|--\right\rangle\}$, the Hamiltonian $H_S$ can be written in the following matrix form: 
\begin{equation}
H_{S}=\left(\begin{array}{cccc}
2B & 0 & 0 & 0\\
0 & 0 & J & 0\\
0 & J & 0 & 0\\
0 & 0 & 0 & -2B
\end{array}\right)\;.
\end{equation}
The eigenvalues of the above Hamiltonian $H_S$ are given by
\begin{equation}
\label{ev}
E_{1}=-2B,\, E_{2}=2B,\, E_{3}=-J,\, E_{4}=+J\,,
\end{equation}
 with the respective eigenstates
 \begin{eqnarray}
&& |E_{1}\rangle=\left(\begin{array}{c}
0\\
0\\
0\\
1
\end{array}\right)=|--\rangle\;,\;|E_{2}\rangle=\left(\begin{array}{c}
1\\
0\\
0\\
0
\end{array}\right)=|++\rangle\;,\nonumber\\
&&|E_{3}\rangle=\frac{1}{\sqrt{2}}\left(\begin{array}{c}
0\\
-1\\
1\\
0
\end{array}\right) = \frac{1}{\sqrt{2}}\left(\left|-+\right\rangle -\left|+-\right\rangle \right)\;,\nonumber\\
&&\; |E_{4}\rangle=\frac{1}{\sqrt{2}}\left(\begin{array}{c}
0\\
1\\
1\\
0
\end{array}\right)=\frac{1}{\sqrt{2}}\left(\left|-+\right\rangle +\left|+-\right\rangle \right)\;.
\label{eigen}
 \end{eqnarray}

Further, within the Lamb-Dicke limit, it is assumed that the ionic vibration is confined to its two lowest lying energy levels, while the higher excited states are not populated. As a result, the vibrational mode can be considered as a two-level cold bath (with the relevant phonon-number states $|0\rangle$ and $|1\rangle$) with an average phonon number $\bar{n}_{\rm ph}\ll 1$. For example, one can achieve $\bar{n}_{\rm ph}\approx 0.02$ in a single Be ion, that can be cooled using standard ion trapping technique \cite{Monroe ion trap}. We emphasize that a finite-level system can act as a bath, as coupling to such bath often leads to decoherence of the system (see, e.g., \cite{Mam and Brummer paper}). Here the system S continuously interacts with this effective cold bath through the Hamiltonian $H_{\rm int}$, while the thermal environment at an equilibrium temperature $T_H$  interacts with both the system S and the vibrational mode.

%

\subsection{Implementation of different strokes}

In the following, we focus on the quantum Otto cycle, that consists of four strokes: two isochoric strokes and two adiabatic strokes.  Here we show how to implement these strokes with the system S and the two baths as identified above.

\begin{figure}
\includegraphics[width=8cm,height=6cm]{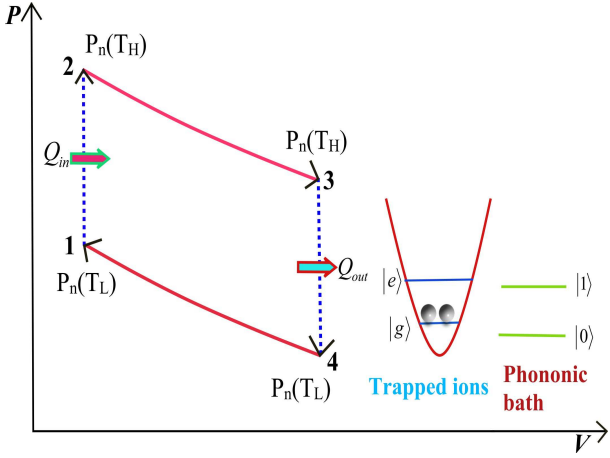}
\caption{Schematic diagram of a quantum Otto cycle using two trapped ions. The solid red (dashed blue) lines refer to the adiabatic (isochoric) processes. The insets display the relevant electronic states and the states of the vibrational mode.}
\end{figure}

\subsubsection{Ignition stroke}
In this isochoric process ($1\rightarrow 2$, see Fig. 1), the ions interacts with the hot bath and get thermalized to an equilibrium temperature $T_H$. To estimate the heat exchanged by the system with the bath during this stroke, we start by rewriting the Hamiltonian $H_1$ in the joint basis 
$\{\left|++1\right\rangle ,\left|++0\right\rangle ,\left|+-1\right\rangle ,\left|-+1\right\rangle ,\left|--1\right\rangle ,\left|+-0\right\rangle ,$ $\left|-+0\right\rangle ,\left|--0\right\rangle \}$ in the following matrix form:
\begin{equation}
H_1=\left(\begin{array}{cccccccc}
2B+\omega & 0 & 0 & 0 & 0 & 0 & 0 & 0\\
0 & 2B & k & k & 0 & 0 & 0 & 0\\
0 & k & \omega & J & 0 & 0 & 0 & 0\\
0 & k & J & \omega & 0 & 0 & 0 & 0\\
0 & 0 & 0 & 0 & -2B+\omega & k & k & 0\\
0 & 0 & 0 & 0 & k & 0 & J & 0\\
0 & 0 & 0 & 0 & k & J & 0 & 0\\
0 & 0 & 0 & 0 & 0 & 0 & 0 & -2B
\end{array}\right)\;,
\end{equation}
where we have assumed that coupling between the electronic states and the vibrational mode are the same for both the ions, i.e., $k_{1}=k_{2}=k$. The eigenstates $|U_n\rangle$ may be written in terms of the joint basis  as
\begin{eqnarray}
\left|U_{n}\right\rangle  & = & a_{1n}\left| ++1\right\rangle+a_{2n}\left| ++0\right\rangle+a_{3n}\left| +-1\right\rangle\nonumber \\
& & +a_{4n}\left| -+1\right\rangle+a_{5n}\left| --1\right\rangle+a_{6n}\left| +-0\right\rangle\nonumber \\
& & +a_{7n}\left| -+0\right\rangle+a_{8n}\left| --0\right\rangle, \;\;\; n\in [1,8]\;.
\end{eqnarray}

The interaction with the thermal bath leads to the following mixed state:
\begin{equation}
\rho_1^{(H)}=\sum_{n=1}^8p_n|U_n\rangle\langle U_n|\;,\;p_n=\frac{\exp\left(-U_n/k_BT_H\right)}{\sum_{n=1}^8\exp\left(-U_n/k_BT_H\right)}\;,
\label{rho1H}
\end{equation}
where $p_n$ is the occupation probability of the $n$th eigenstate $|U_n\rangle$ (with corresponding eigenvalue $U_n$) of the total Hamiltonian $H_1$. Note that this state is achieved at the steady state irrespective of the initial preparation of the ions.

The reduced density matrix of the system S can be obtained by taking the partial trace over the vibrational states as
\begin{eqnarray}
\rho_{S}^{(H)} & = & \frac{1}{P_{H}}\sum_{n=1}^{8}e^{-U_{n}/K_{B}T_{H}}\left[\left|++\right\rangle \left\langle ++\right|(a_{1n}^{2}+a_{2n}^{2})\right. \nonumber \\
& & +\left|+-\right\rangle \left\langle +-\right|(a_{3n}^{2}+a_{6n}^{2}) +\left|-+\right\rangle\left\langle -+\right|\nonumber(a_{4n}^{2}+a_{7n}^{2})\nonumber \\
 & &+\left|--\right\rangle \left\langle --\right|(a_{5n}^{2} +a_{8n}^{2})\nonumber \\
  & & +(a_{1n}a_{3n} +a_{2n}a_{6n}) (\left|++\right\rangle \left\langle +-\right|+\left| +-\right\rangle \left\langle ++\right|) \nonumber \\
  & &+(a_{1n}a_{4n} +a_{2n}a_{7n}) (\left| ++\right\rangle \left\langle -+\right| + \left| -+\right\rangle  \left\langle ++\right|)\nonumber \\ 
  & &+(a_{1n}a_{5n}+a_{2n}a_{8n})(\left|++\right\rangle \left\langle --\right|+\left| --\right\rangle\left\langle ++\right|)\nonumber \\
  & &  +(a_{3n}a_{4n}+a_{6n} a_{7n})(\left| +-\right\rangle \left\langle -+\right|+\left| -+\right\rangle\left\langle +-\right|)\nonumber \\
  & &  +(a_{3n}a_{5n}+a_{6n}a_{8n})(\left|+-\right\rangle \left\langle --\right| +\left| --\right\rangle  \left\langle +-\right|)\nonumber \\
  & &\left.+(a_{4n}a_{5n}+a_{7n}a_{8n}) (\left|-+\right\rangle \left\langle --\right|+\left| --\right\rangle\left\langle -+\right|)\right]
\end{eqnarray}
where $P_H=\sum_{n=1}^8 \exp[-U_n/k_BT_H]$ is the normalization constant.

This can be rewritten in terms of the energy eigenstates $|E_i\rangle$ of the system Hamiltonian $H_S$ through the inverse transformation of the Eq. (\ref{eigen})
\begin{eqnarray}
\left|++\right\rangle =\left|E_{2}\right\rangle \;,\left|+-\right\rangle =\frac{1}{\sqrt{2}}\left(\left|E_{4}\right\rangle -\left|E_{3}\right\rangle \right)\;, \nonumber \\
\left|-+\right\rangle =\frac{1}{\sqrt{2}}\left(\left|E_{4}\right\rangle +\left|E_{3}\right\rangle \right)\;, 
 \left|--\right\rangle =\left|E_{1}\right\rangle \;.
\end{eqnarray}
Using above equation, we can get the occupation probability $P_{i}$ of the $i$th 
eigenstate $\left|E_{i}\right\rangle$ ($i\in [1,4]$) as follows:
\begin{eqnarray}
P_{1}\left(T_{H}\right)&=&\sum_{n=1}^{8}e^{-U_{n}/K_{B}T_{H}}\left(a_{5n}^{2}+a_{8n}^{2}\right)\;,\nonumber\\
P_{2}\left(T_{H}\right)&=&\sum_{n=1}^{8}e^{-U_{n}/K_{B}T_{H}}\left(a_{1n}^{2}+a_{2n}^{2}\right)\;,\nonumber\\
P_{3,4}\left(T_{H}\right)&=& \sum_{n=1}^{8}e^{-U_{n}/K_{B}T_{H}}\Big[\Big(\frac{a_{3n}^{2}\mp a_{3n}a_{4n}}{2}\Big)+\nonumber \\&&\Big(\frac{a_{4n}^{2}\mp a_{3n}a_{4n}}{2}\Big)
+\Big(\frac{a_{6n}^{2}\mp a_{6n}a_{7n}}{2}\Big)\nonumber \\&&+\Big(\frac{a_{7n}^{2}\mp a_{6n}a_{7n}}{2}\Big)\Big]\;.
\end{eqnarray}

The average energy of the system under consideration can be written as $U=\sum_{i=1}^4 E_i P_i$. Here the change in the $E_i$s corresponds to the heat exchange, while the change in the probabilities refer to the certain work done during the cycle \cite{Kieu}. Based on the initial preparation of the ion, if the initial probability for being in the $i$th eigenstate is $P_{i}\left(T_{L}\right)$, then the heat exchanged with the hot bath  by the system S during this stroke is given by
\begin{equation}
Q_{H}=\sum_{i=1}^{4}E_{i}^{H}\left\{ P_{i}\left(T_{H}\right)-P_{i}\left(T_{L}\right)\right\}\;. 
\end{equation}
Note that in this process, the magnetic field is kept fixed at $B=B_H$, such that the eigenvalues $E_{i}^{H}$ of the system Hamiltonian $H_S$ also remain constant and therefore no work is done. Due to the change in the occupation probabilities, only the heat is exchanged during this cycle. 

\subsubsection{Expansion stroke}
During this adiabatic cycle ($2\rightarrow 3$, see Fig. 1), the magnetic field is modified from $B_H$ to $B_L$, such that the occupation probabilities of the four eigenstates 
$\left\{\left|E_{i}\right\rangle, i=1,2,3,4\right\}  $ remain unchanged. Consequently there is no heat exchange between the system and heat bath. However, the corresponding eigenvalues $E_{i}^{H}$ change to the values $E_{1,2}^{L}=\mp2B_{L}$ and $E_{3,4}^{L}=\mp J$. This amounts to the following work done by the system S during this cycle:
\begin{equation}
W_{1}=\sum_{i=1}^{4}P_{i}\left(T_{H}\right)\left(E_{i}^{L}-E_{i}^{H}\right)\;.
\end{equation}

\subsubsection{Exhaust stroke}
In an usual Otto engine, this stroke is associated with cooling of the system through heat release to the cold bath. In the present case, in this stroke ($3\rightarrow 4$, see Fig. 1), the system exchanges heat $Q_L$  with the cold bath and the system Hamiltonian changes from $H_{S}\left(B_{H}\right)$
to $H_{S}\left(B_{L}\right)$. To estimate the $Q_L$, we start with the following state of the ions, that is adiabatically evolved thermal state, as attained at the end of the expansion stroke:
\begin{equation}
\rho_1^{(L)}=U_I^\dag\rho_1^{(H)}U_I\;,
\end{equation}
where
\begin{eqnarray}
U_{I}& = &{\cal T} \exp\left[-i\int_{0}^{\tau}dt'H_1\left(t'\right)\right]\;,\nonumber\\
H_1\left(t\right)& = &H_{S}\left(t\right)+H_{\rm ph}+H_{\rm int}\;,\nonumber\\
H_{S}\left(0\right)& = & J\left(\sigma_{+}^{(1)}\sigma_{-}^{(2)}+\sigma_{-}^{(1)}\sigma_{+}^{(2)}\right)+B_H\left(\sigma_{z}^{(1)}+\sigma_{z}^{(2)}\right)\;, \nonumber \\
H_{S}\left(\tau\right)& = &J\left(\sigma_{+}^{(1)}\sigma_{-}^{(2)}+\sigma_{-}^{(1)}\sigma_{+}^{(2)}\right)+B_L\left(\sigma_{z}^{(1)}+\sigma_{z}^{(2)}\right)\;. 
\end{eqnarray}
Here ${\cal T}$ stands for time-ordering and $\tau$ is the finite time for the adiabatic change of the magnetic field from $B_H$ and $B_L$ during the expansion stroke. The state $\rho_{1}^{\left(L\right)}$
can be written in the joint basis of the electronic states and the vibrational mode
as
\begin{eqnarray}
\rho_{1}^{(L)}& = &\sum_{m,n =1}^8\rho_{1}^{\left(mn\right)}\left|m\right\rangle \left\langle n\right|\;\;,\;\;
\left|m\right\rangle , |n\rangle=\left|++1\right\rangle ,\left|++0\right\rangle,\nonumber \\
& & \left|\pm\mp1\right\rangle ,\left|--1\right\rangle ,\left|\pm\mp0\right\rangle ,\left|--0\right\rangle\;.
\end{eqnarray}

Note that, at thermal equilibrium, the system S is entangled with the vibrational mode. To this end, we propose a projective measurement of the state of the system S, thereby disentangling S from the vibrational mode.  Further, if the two-ion system is measured in the ground state, the occupation probabilities of the higher excited states reduce to zero. This mimics the release of heat to the cold bath, as is usually required in an exhaust stroke of an Otto engine. Clearly, such a measurement-induced heat exchange depends upon the choice of the projected state. In fact, as shown later in this Section, the system can work as a quantum heat engine or a quantum refrigerator, depending upon the measurement basis.  In the following, we choose the system eigenstates $|E_i\rangle$ as the measurement basis.  

Generally speaking, upon projection onto the eigenstate $|E_i\rangle$ of the system, the density matrix $\rho_1^{(L)}$ gets factorized and can be written as
\begin{equation}
\rho_1^{(L)}|_{\rm meas}=|E_i\rangle\langle E_i|\sum_{k,l=0}^{1}r_{kl}^{(i)}\left|k\right\rangle \left\langle l\right|\;,
\end{equation}
where $\left|k\right\rangle, k\in 0,1$ represents the vibrational states of the ions
and $r_{kl}^{(i)}$ are the relevant density matrix elements between
the states $\left|k\right\rangle ,\left|l\right\rangle $, corresponding to the projection onto $|E_i\rangle$.
In this way, the system exchanges heat with the cold bath and thereafter gets decoupled from the cold bath.

Through this heat-exchange process, the probability distribution $\{P_i\}$ of the eigenstates also changes, while maintaining the corresponding eigenvalues identical. The local magnetic field $B_{L}$ is kept constant during this stroke. If the final occupation probability for the $i$th eigenstate becomes $P_i(T_L)$,  the heat exchange between the system and the cold bath can be calculated as
\begin{equation}\label{ql1}
Q_L=\sum_{i=1}^4E_i^{L}[P_i(T_L)-P_i(T_H)]\;.
\end{equation}
Note that, as in the ignition stroke, no work is done during this stroke as well.
The measurement process, as described above, is apparently probabilistic and relies on the result of the measurement. A reasonable alternative option for decoupling the system from the bath could be to use the
non-selective measurement, as described in \cite{non-selective measurement}.

\begin{figure}[!h]
$\begin{array}{cc}
 \includegraphics[width=0.49\linewidth]{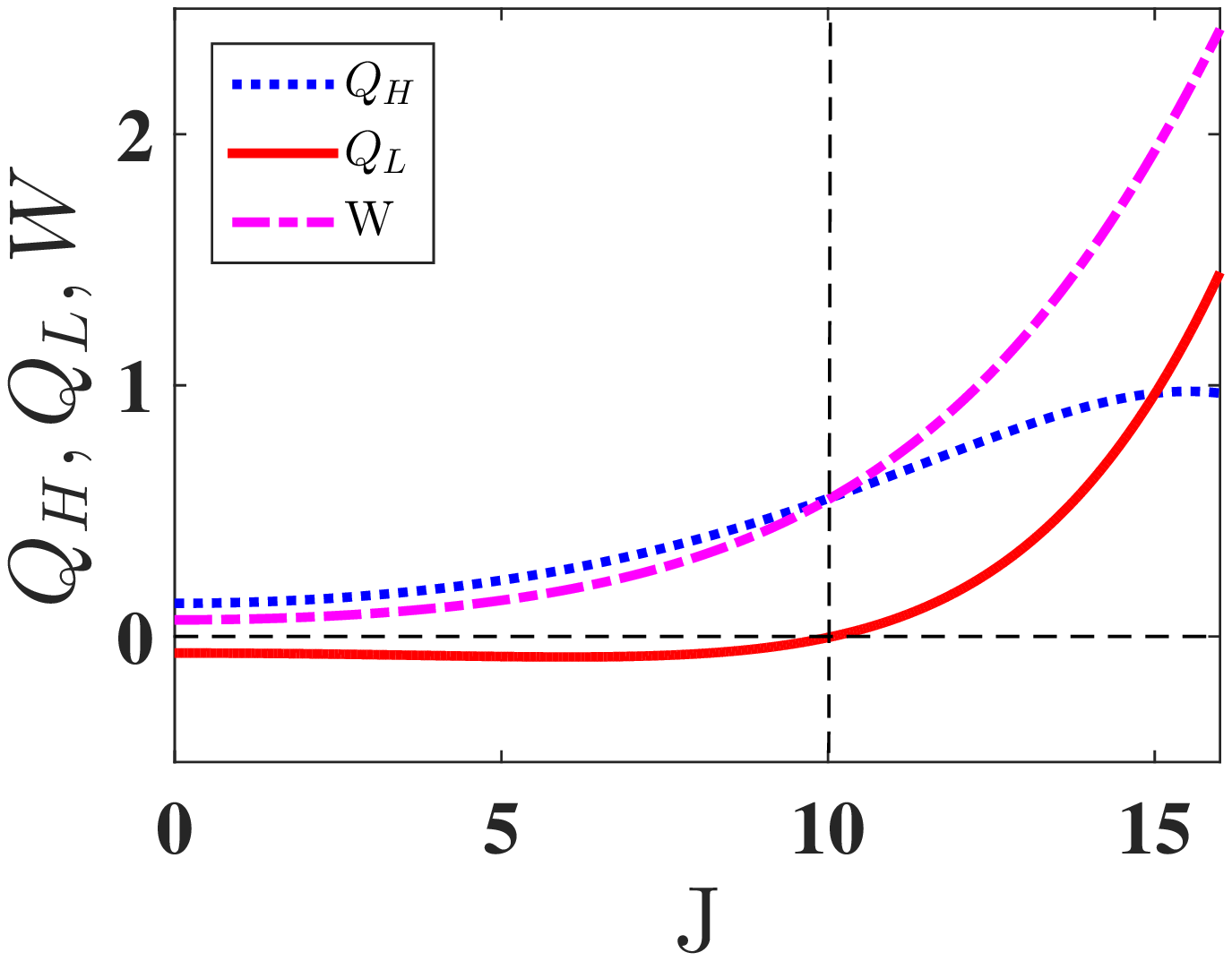}&
\includegraphics[width=0.49\linewidth]{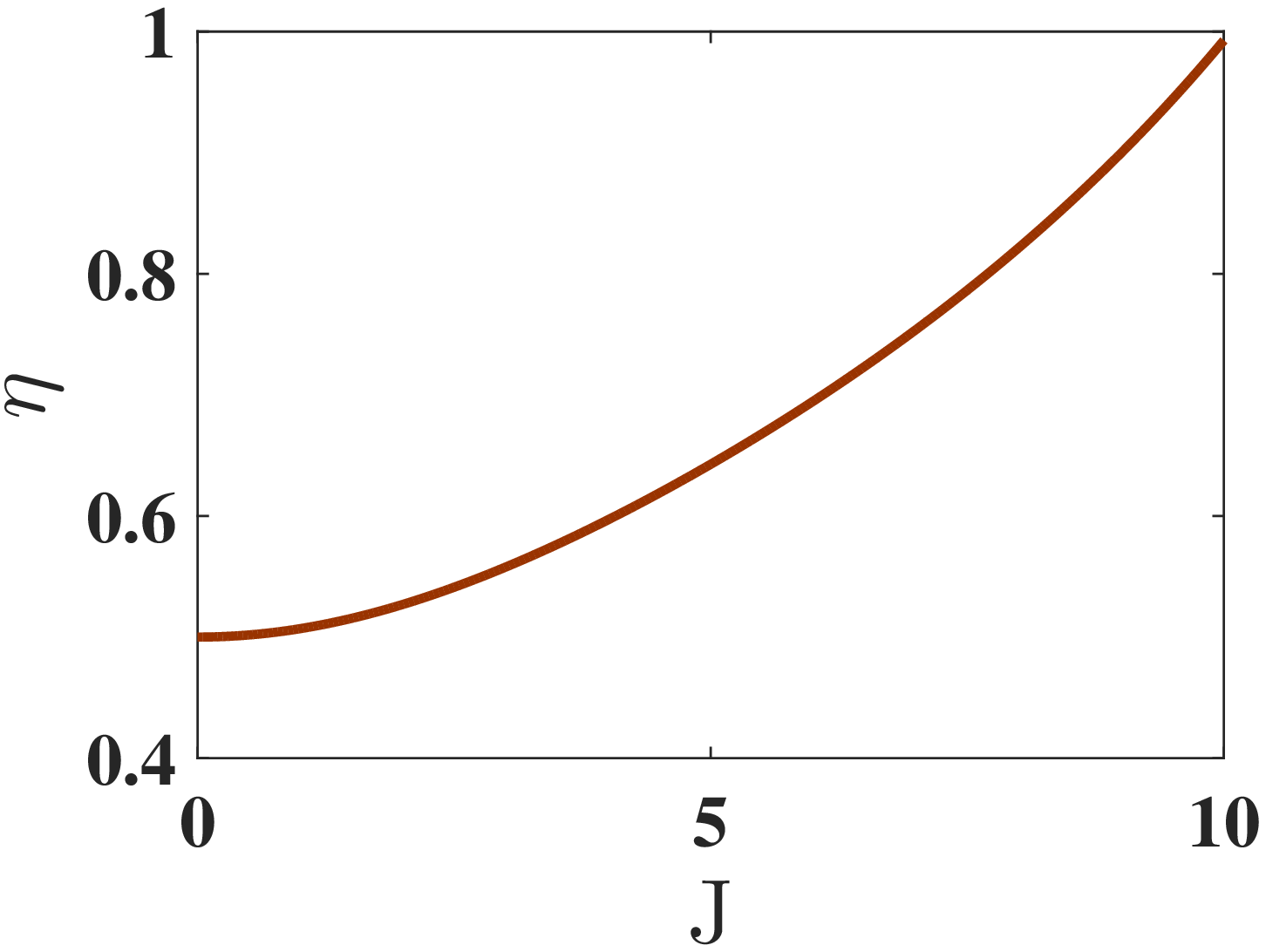}\\(a) & (b)
\end{array}$ 
\caption{ Variation of (a) heat exchanged $Q_H$ (dotted blue) and $Q_L$ (solid red), with the hot and the cold bath, respectively and the net work done (dot-dashed magenta) $W$ and (b) the efficiency $\eta$  as a function of the coupling constant $J$. The others parameters for the cycle are $B_H = 10, B_L = 5, k=0.1, \omega =1, k_BT_H=3.5$. Here the measurement is done in the basis  $|E_1\rangle$. The physically acceptable parameter region for the engine to operate is  $J\le 2B_{L}=10$.
}
\end{figure}
\subsubsection{Compression Stroke}
For this stroke ($4\rightarrow 1$), the system goes through an adiabatic evolution once more, during which the magnetic field strength is adiabatically changed from $B_L$ to $B_H$. The system remains in contact with the hot bath. During the expansion stroke, the occupation probabilities of the energy eigenstates $|E_i\rangle$ remain unaltered. The eigenvalues change from $E_i^{L}$ to $E_i^{H}$ due to the change in the magnetic field. This leads to the following work done during this stroke:
\begin{equation}
W_2=\sum_{i=1}^4P_i(T_L)(E_i^{H}-E_i^{L})\;.
\end{equation}
It must be borne in mind that after the compression stroke ends, the heat machine goes into the next cycle, starting with the ignition stroke. During this stroke, the system gets thermalized to the state (\ref{rho1H}), irrespective of its initial state and therefore the cycle continues in a similar fashion. Further, as long as the Lamb-Dicke limit is maintained, the vibrational mode remains confined to its two lowest energy levels and can be reused as a cold bath during the next cycle. 

\begin{figure}[!h]
$\begin{array}{cc}
\includegraphics[width=0.49\linewidth]{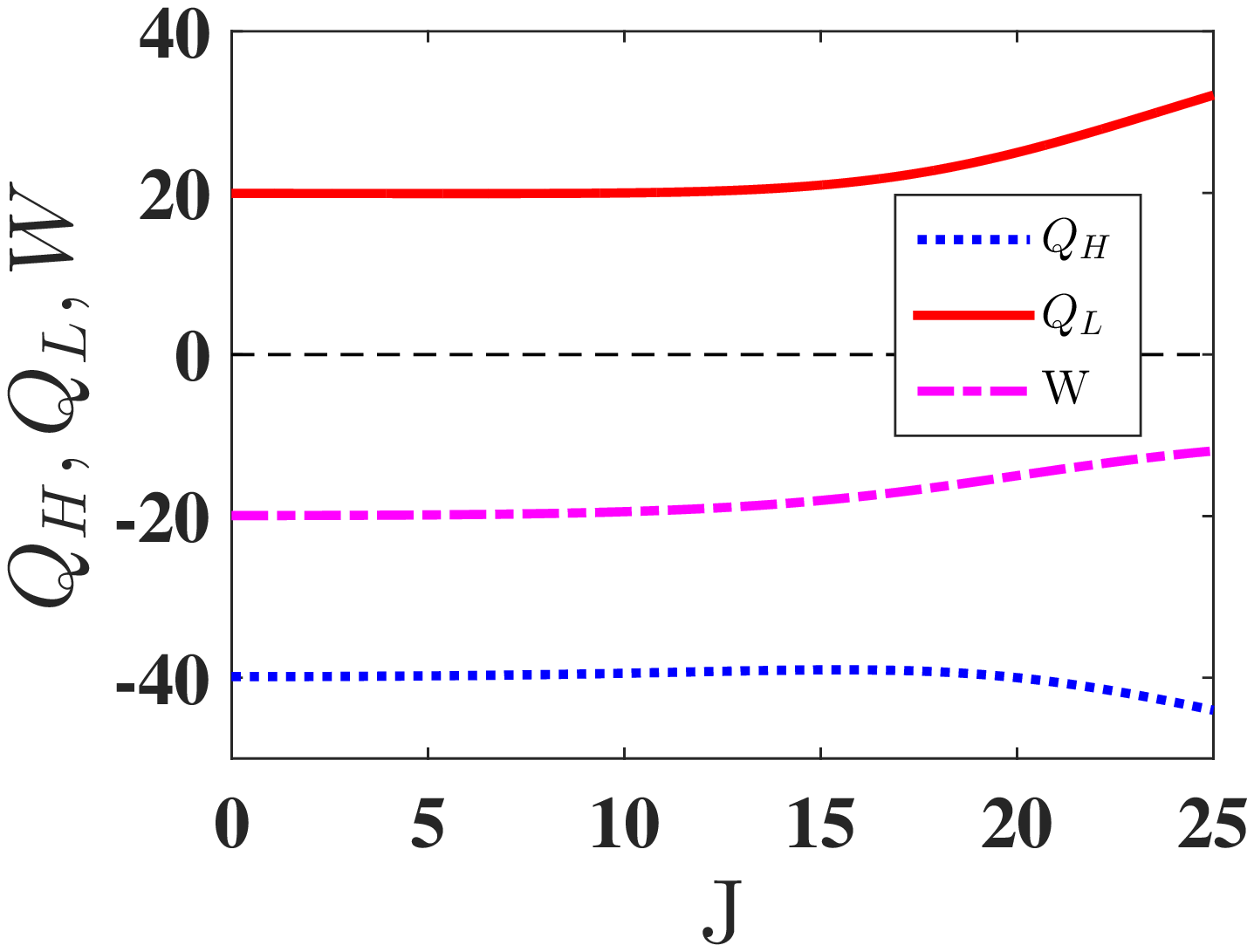}&
\includegraphics[width=0.49\linewidth]{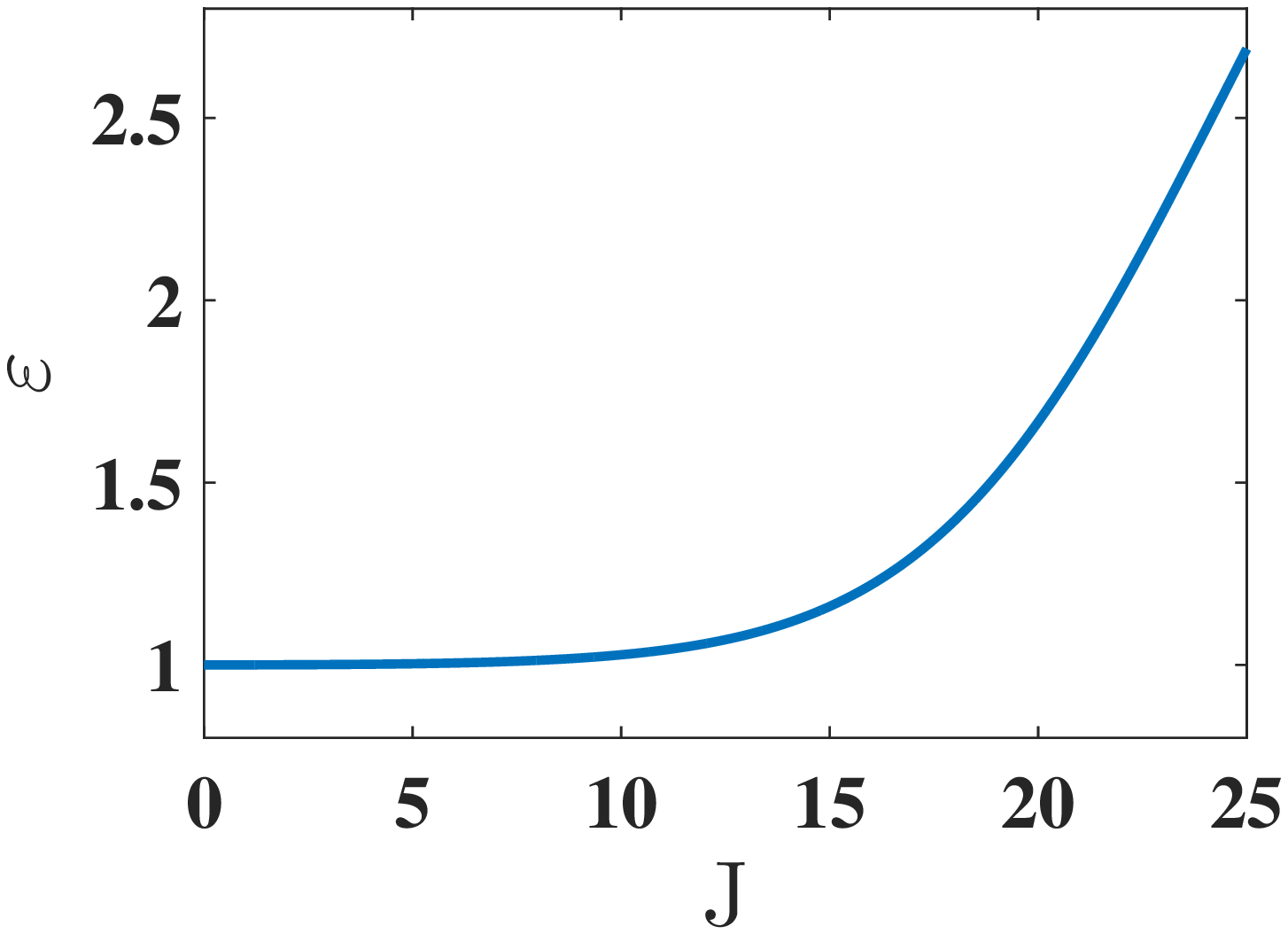}\\(a) & (b)
\end{array}$ 
\caption{Variation of (a) heat exchanged $Q_H$(dotted blue) and $Q_L$(solid red), with the hot and the cold bath, respectively and the net work done $W$ (dot-dashed magenta) and (b) the COP $\varepsilon$  as a function of the coupling constant $J$. The parameters are the same as  in Fig. 2. Here the measurement is done on the state $|E_2\rangle$.}
\end{figure}

\subsection{Efficiency and COP of the heat machine}
In the following, we consider the measurement in different eigenstates of $H_S$. 

{\it Case I: Projection in $|E_1\rangle$ state\/}:
This state is the ground state as long as $J<2B$ [see Eq. (\ref{ev})]. So the projection of the state of the system S into $|E_1\rangle$ corresponds to cooling of the system. This heat can be thereby extracted from the system and transferred into the vibratioanal mode. 
In this case, we find that the heat released into the cold bath is $Q_L<0$, while the heat absorbed by the system becomes $Q_H>0$ [see Fig. 2(a)].  Further, the system does certain work during the two adiabatic strokes, such that the total work done $W>0$. This situation clearly refers to executing a quantum heat engine.
\begin{figure}[!h]
$\begin{array}{cc}
\includegraphics[width=0.49\linewidth]{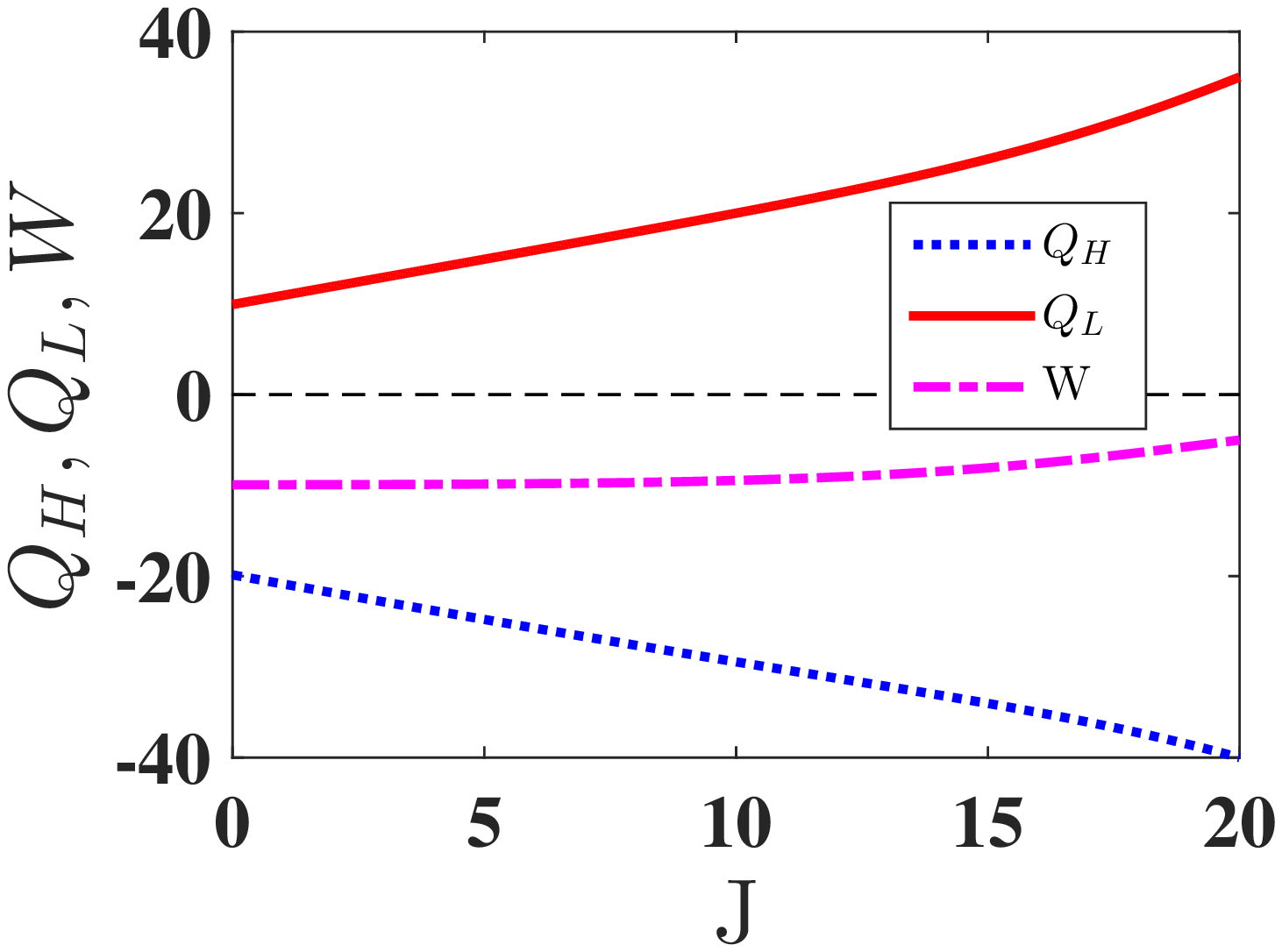}&
\includegraphics[width=0.49\linewidth]{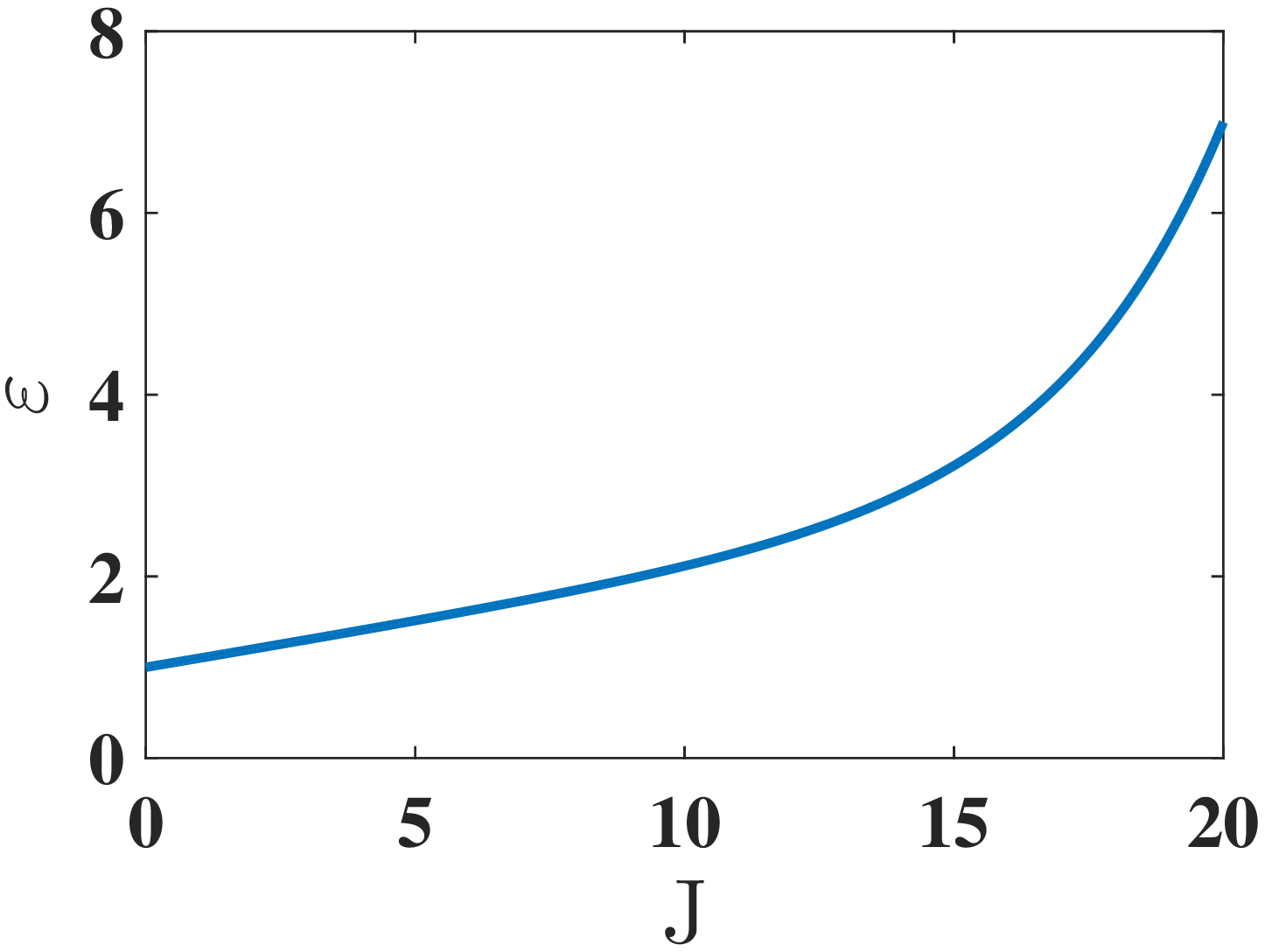}\\(a) & (b)
\end{array}$ 
\caption{Variation of (a) heat exchanged $Q_H$ (dotted blue) and $Q_L$ (solid red), with the hot and the cold bath, respectively and the net work done $W$ (dot-dashed magenta) and (b) the COP $\varepsilon$  as a function of the coupling constant $J$. The parameters are the same as  in Fig. 2. Here the measurement is done on the state $|E_4\rangle$.}
\end{figure}
The efficiency of the heat engine is defined as $\eta=\frac{\rm Work\,Output}{\rm Heat\,Input}=\frac{Q_{H}+Q_{L}}{Q_{H}}$. 
It is easy to see from the Fig. 2(a)  that the system S behaves like a heat engine for the parameter regime $J\le2B_{L}$, i.e., as long as $\left|E_{1}\right\rangle $ remains the ground state. Note that in this regime, the efficiency increases for increasing values of $J$ and becomes near to unity. Beyond this regime, one attains a unphysical situation. In Fig. 2(b), we show how the efficiency $\eta$ approaches unity with increase in $J$ to its upper limit $2B_L$. Further in the limiting case of uncoupled spins (i.e., $J=0$), the efficiency becomes $\eta_{0}=1-\frac{B_{L}}{B_{H}}$, that matches with the results for a single-spin quantum Otto engine \cite{Our Paper}. 

{\it Case II: Projection in $|E_2\rangle$ and $|E_4\rangle$ states\/}: If the measurement is done in the other eigenstates, we obtain a possibility of the refrigerator action, in which the system absorbs heat from the cold bath ($Q_{L}>0$) and releases heat into the hot bath ($Q_{H}<0$). In this process, a certain amount of work is done on the system ($W<0$). In Figs. 3(a) and 4(a), we show that such a situation is obtained over a large range of the coupling constant $J$ for a given set of values of magnetic fields $B_L$ and $B_H$.

\begin{figure}[!h]
$\begin{array}{cc}
\includegraphics[width=0.49\linewidth]{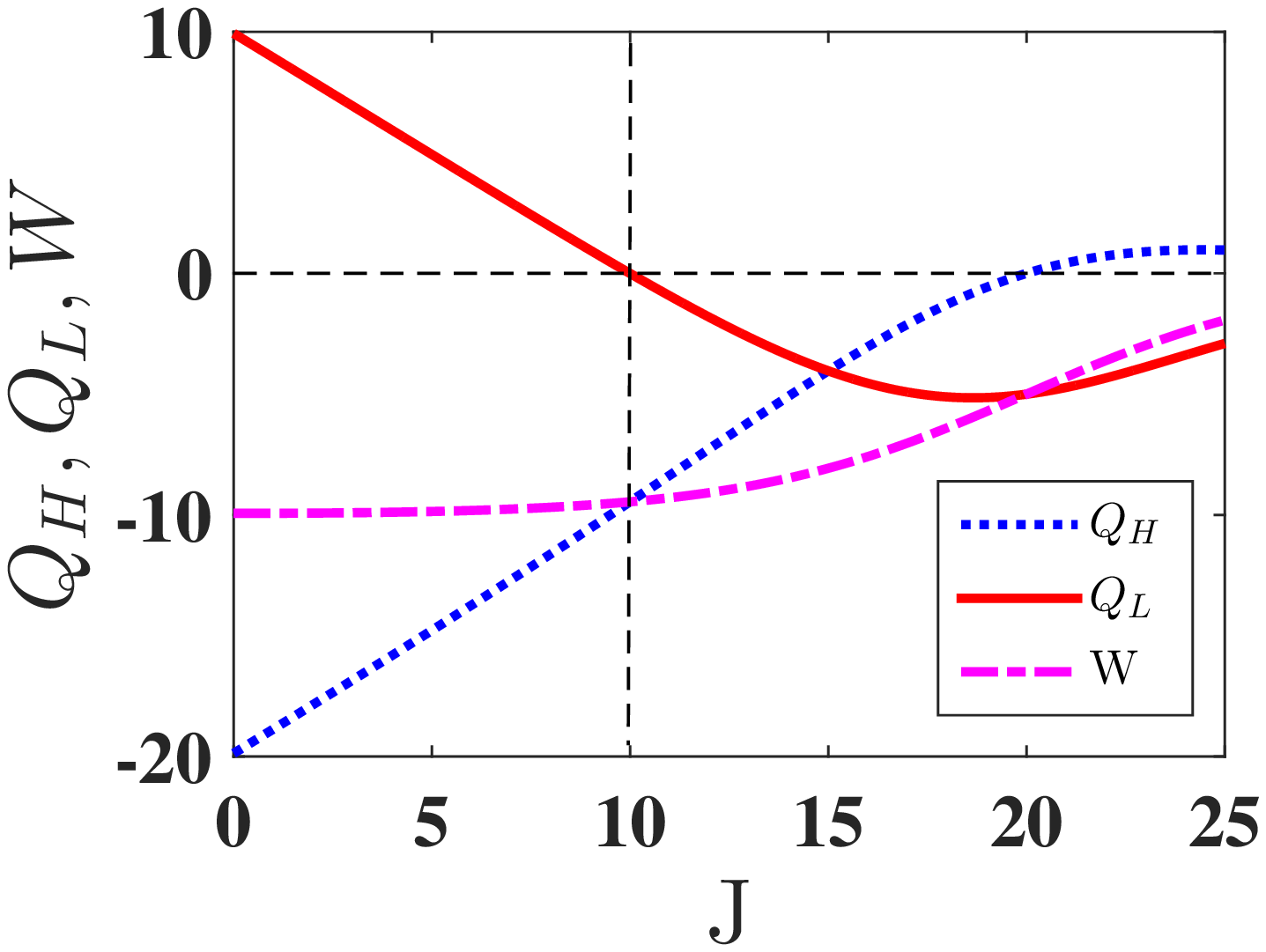}&
\includegraphics[width=0.49\linewidth]{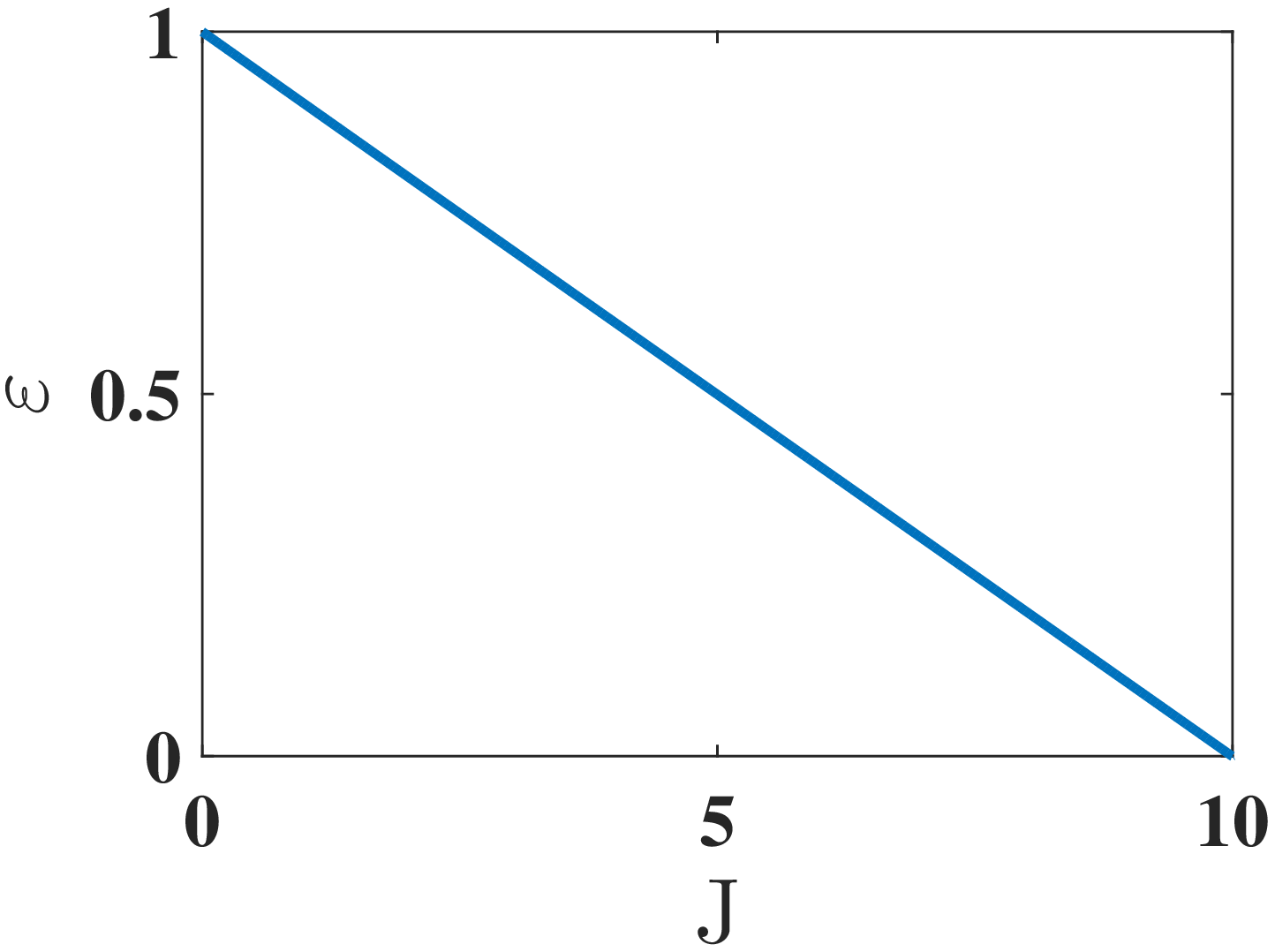}\\(a) & (b)
\end{array}$ 
\caption{Variation of (a) heat exchanged $Q_H$ (dotted blue) and $Q_L$(solid red), with the hot and the cold bath, respectively and the net work done $W$ (dot-dashed magenta) and (b) the COP $\varepsilon$  as a function of the coupling constant $J$. The parameters are the same as  in Fig. 2. Here the measurement is done in the state $|E_3\rangle$. The physically acceptable parameter region for the refrigerator to operate is $J\le 2B_{L}=10$.}
\end{figure}
 
We here emphasize that the states $|E_2\rangle$ and $|E_4\rangle$ are the eigenstates with positive eigenvalues and correspond to excited states. So projecting the system into these eigenstates refers to heating of the system, as one would require in the stroke associated with the cold bath in a refrigeration cycle. The performance of a refrigerator is quantified in terms of the coefficient of performance (COP) $\varepsilon=\frac{\rm Heat\, Input}{\rm |Work\, Output|}=\frac{Q_{L}}{\left|Q_{H}+Q_{L}\right|}$. In Figs. 3(b) and 4(b), we show the variation of $\varepsilon$ with $J$. Clearly the measurement in the $|E_4\rangle$ state leads to a much better performance as a refrigerator for a given value of $J$. 

{\it Case III: Projection in $|E_3\rangle$ state\/}:
Quite interestingly, for a certain regime, $J\le 2B_L$, the measurement in the $|E_3\rangle$  state leads to a refrigeration effect. This is because, in this parameter regime, $|E_3\rangle$ remains an excited state [see Eq. (\ref{ev})] and the measurement in such a state leads to an effective heating of the system S (i.e., $Q_L>0$), while both the $Q_H$ and $W$ remain negative [see Fig. 5(a)]. This refers to a situation, in which the system behaves as a refrigerator. However, for larger values of $J$, the  the performance $\varepsilon$ decreases with $J$, as shown in Fig. 5(b).   For $J>2B_L$, one reaches an unphysical regime, in which neither a heat engine nor a refrigerator action is achievable. 

We show in the parametric plots in Fig. 6 how the efficiency $\eta$ and the COP $\varepsilon$ vary with the work done by the system or on the system, respectively. It is clear from Fig. 6(a) that both the work output and the efficiency of the heat engine are large for $J=2B_L$, if one measures the system in the state $|E_1\rangle$. Similarly, for the measurement in $|E_2\rangle$ [Fig. 6(b)] and $|E_4\rangle$ [Fig. 6(d)] states, the required work to be done becomes less, while the coefficient of performance of the refrigerator increases, as $J$ is increased.  The measurement in $|E_3\rangle$ state is not a desirable choice for refrigeration, because to obtain a large COP, one would require a large amount work [Fig. 6(c)]. 


\begin{figure}[!h]
$ \begin{array}{cc}
\includegraphics[width=0.49\linewidth]{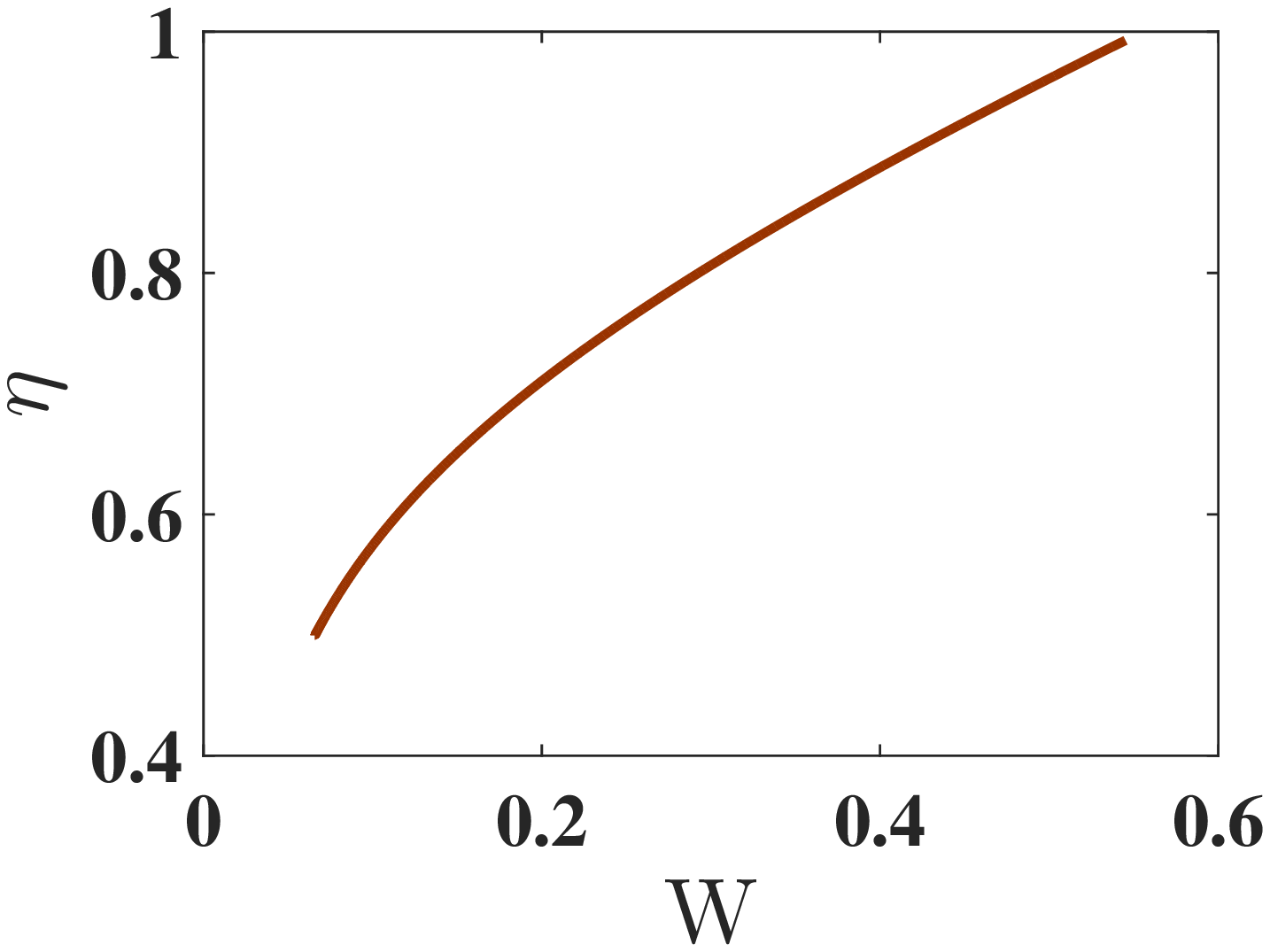}&
\includegraphics[width=0.49\linewidth]{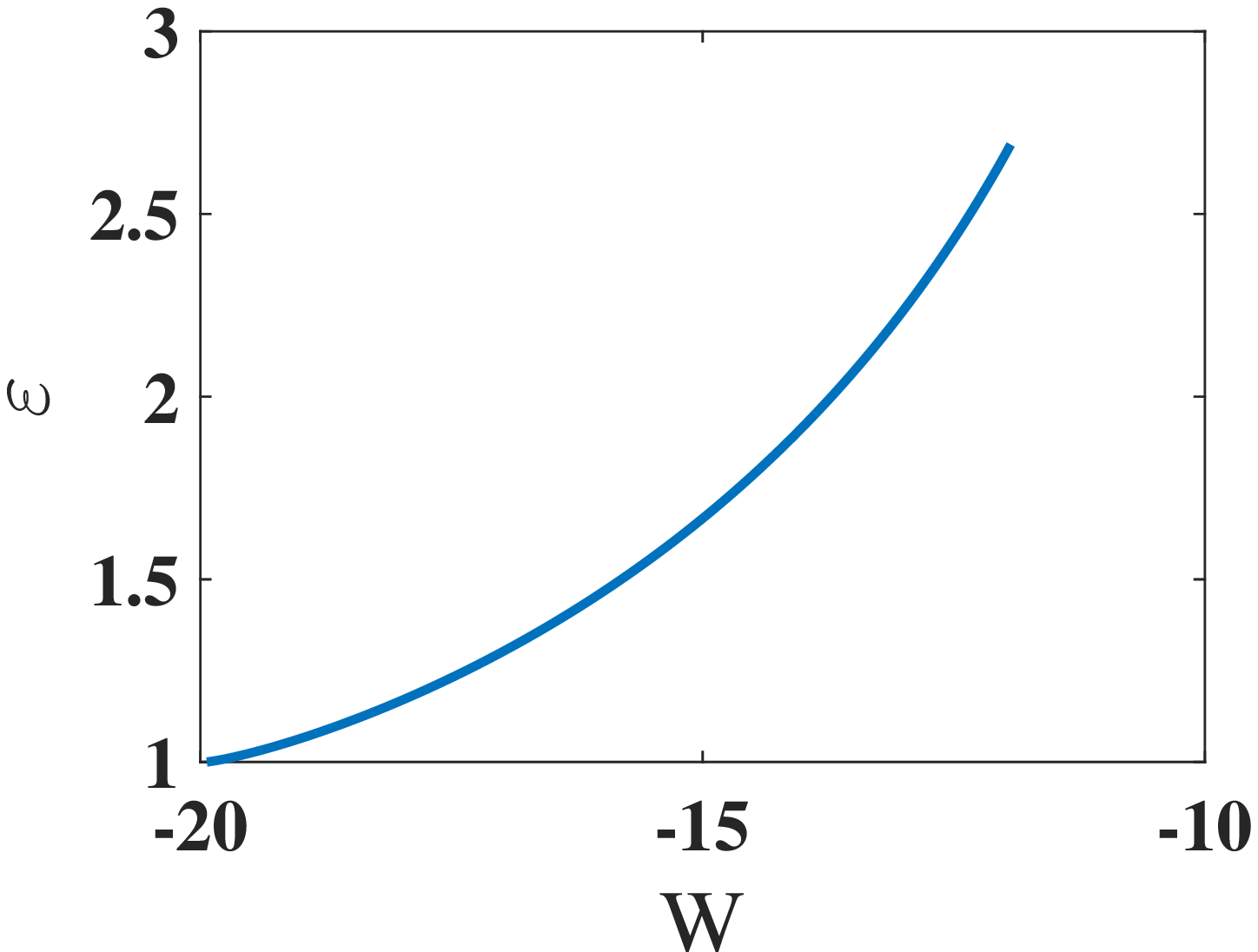}\\(a) & (b)
\end{array} $
$ \begin{array}{cc}
\includegraphics[width=0.49\linewidth]{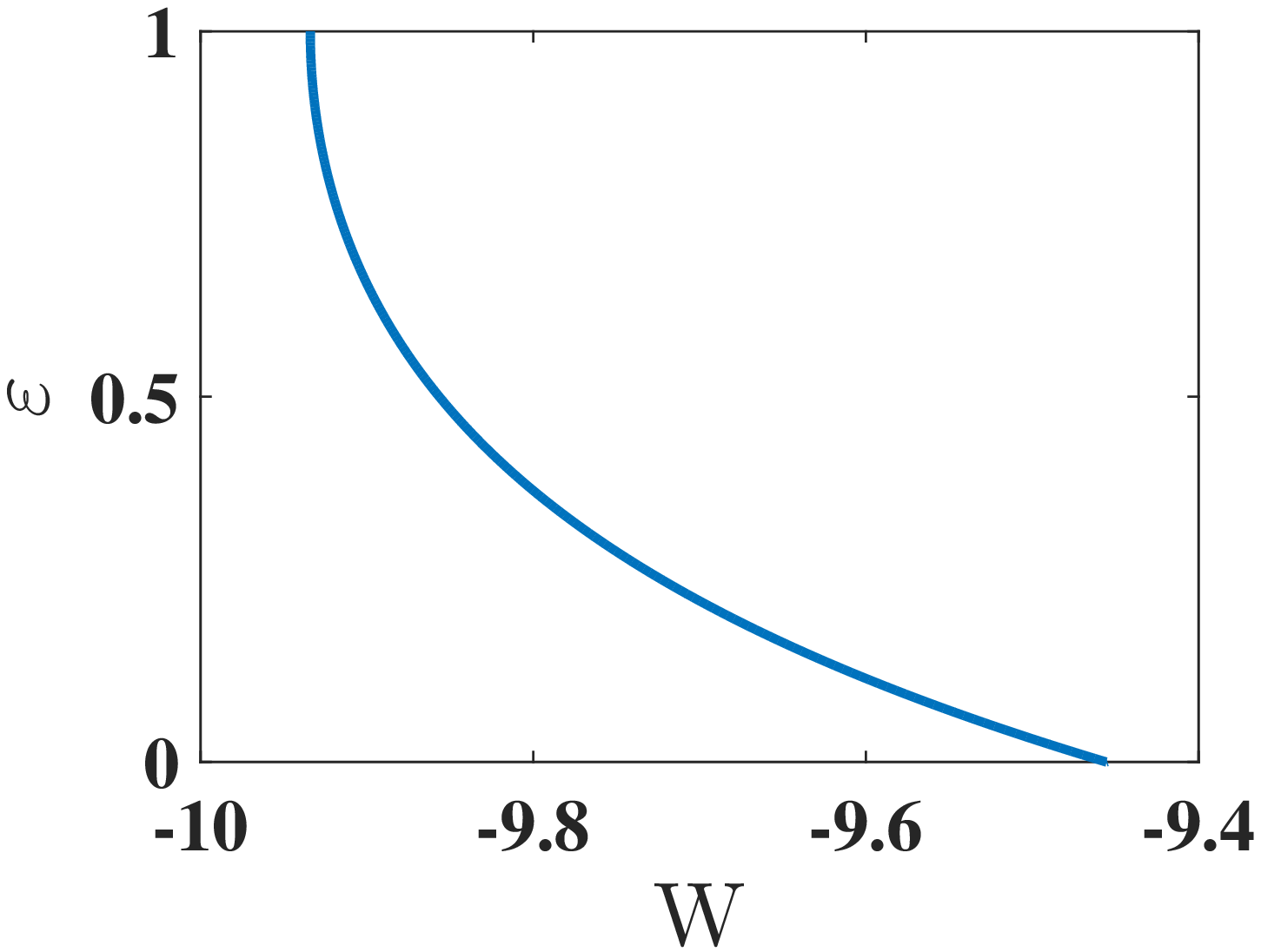}&
\includegraphics[width=0.49\linewidth]{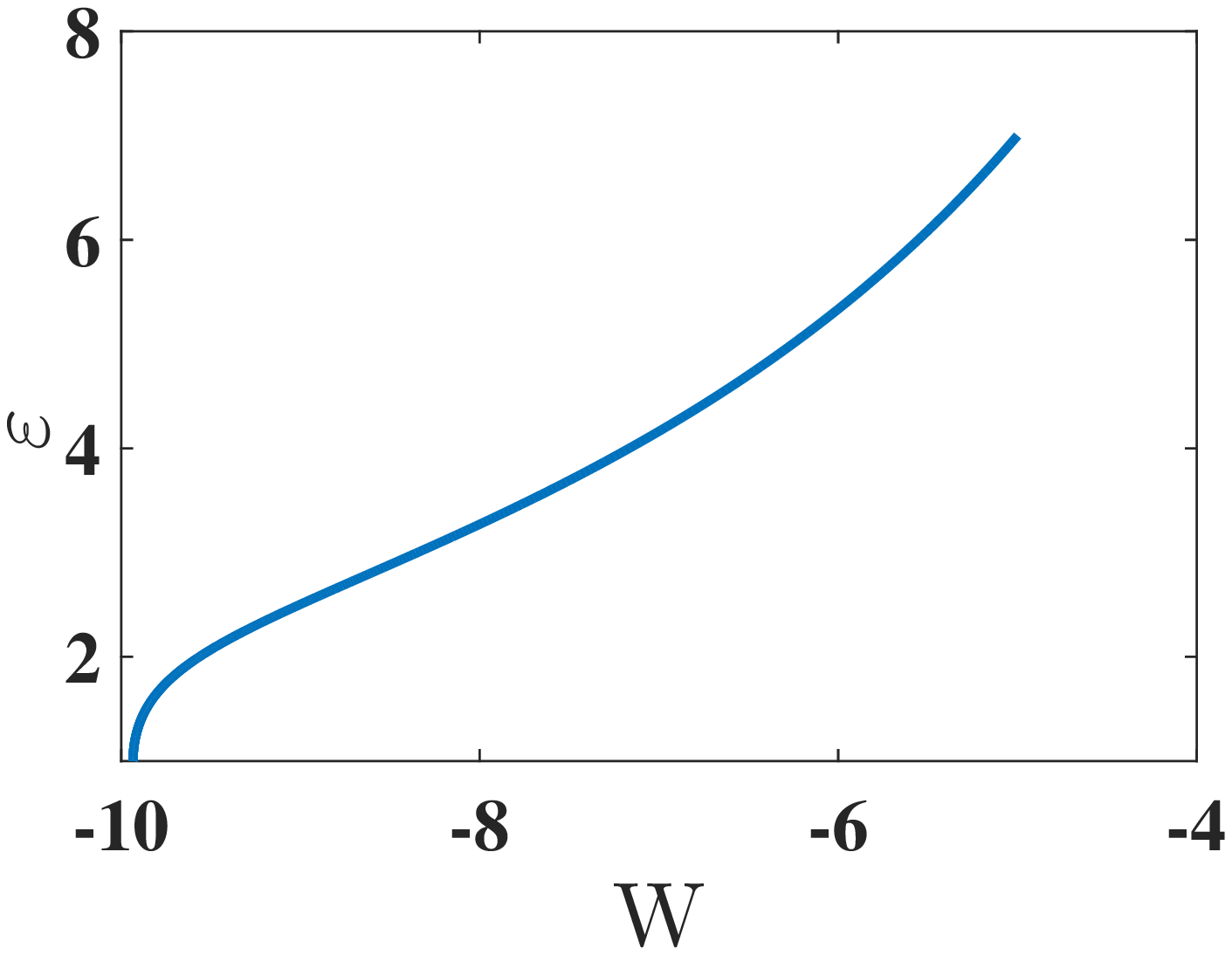}\\(c) & (d)
\end{array} $
\caption{(a) (Color online) Parametric variation of the efficiency $\eta$  with the work done $W$ by the system, when the system is measured in $|E_1\rangle$ state.   Parametric variation of the COP $\varepsilon$ with the work done $W$ on the system, when the system is measured in (b) $|E_2\rangle$ (c) $|E_3\rangle$, and (d) $|E_4\rangle$ states. In all the plots, $J$ is varied from 0 to $2B_{L}$. 
The others parameters are the same as in Fig. 2.
}
\end{figure}
\subsection{Effect of measurement}
In the discussion above, we have not included the measurement cost. This cost would eventually restrict the performance of a heat engine and a refrigerator. The cost for the projective measurement for one qubit is $k_{B}T\ln 2$ \cite{measurement cost}, which is the same as the cost of classical measurement of one bit.  
This leads to modified definition of the efficiency \cite{Abah and Lutz new paper} as
\begin{equation}
\eta'=\frac{\rm energy\,output}{\rm energy\,input}=\frac{Q_{in}+Q_{out}}{Q_{in}+M}\;,
\end{equation}
where $M$ is the cost of measurement. In the present case of two qubits, $M=2 k_{B}T \ln 2$.

\begin{figure}[!h]
$ \begin{array}{cc}
\includegraphics[width=0.49\linewidth]{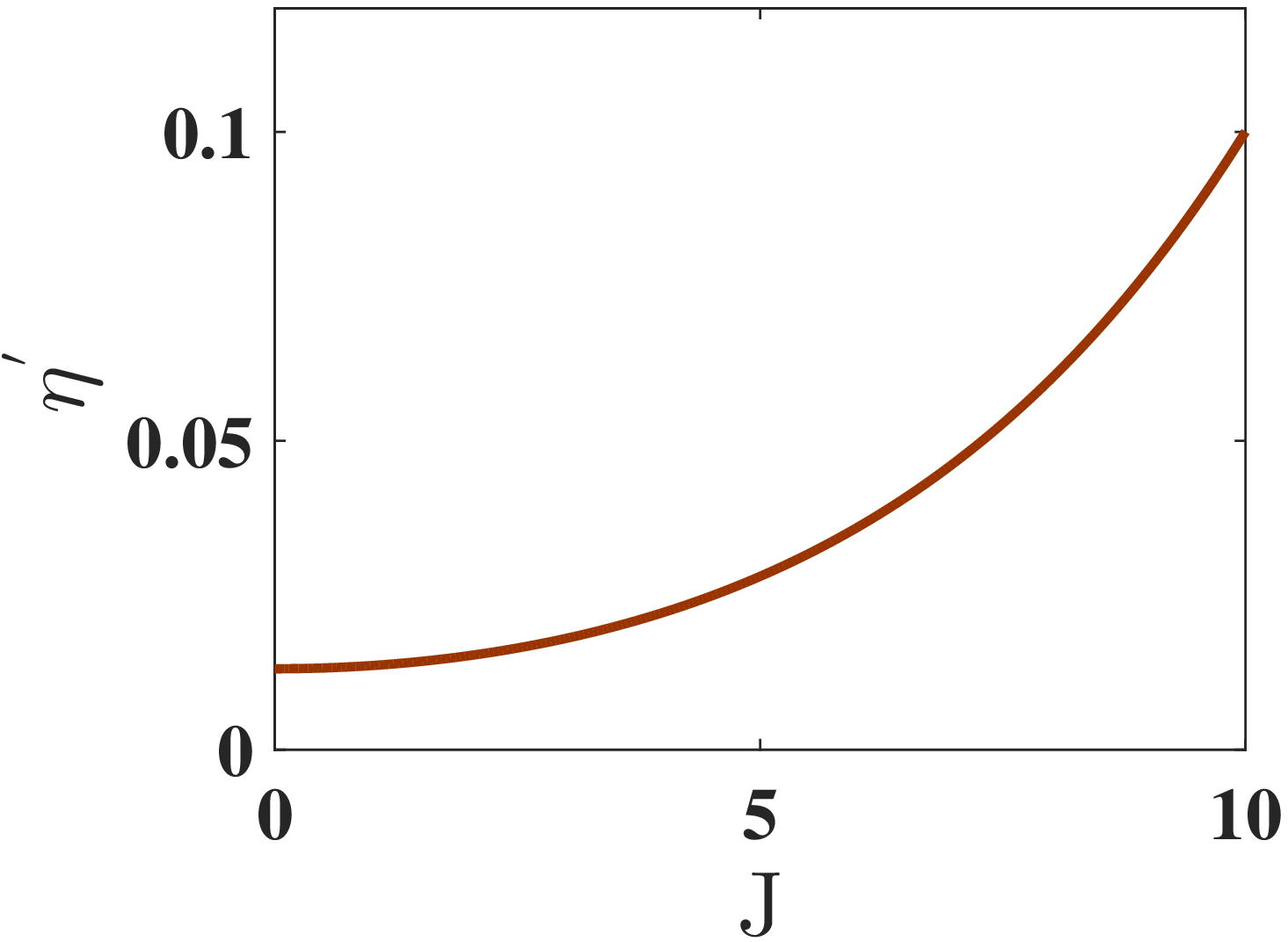}&
\includegraphics[width=0.49\linewidth]{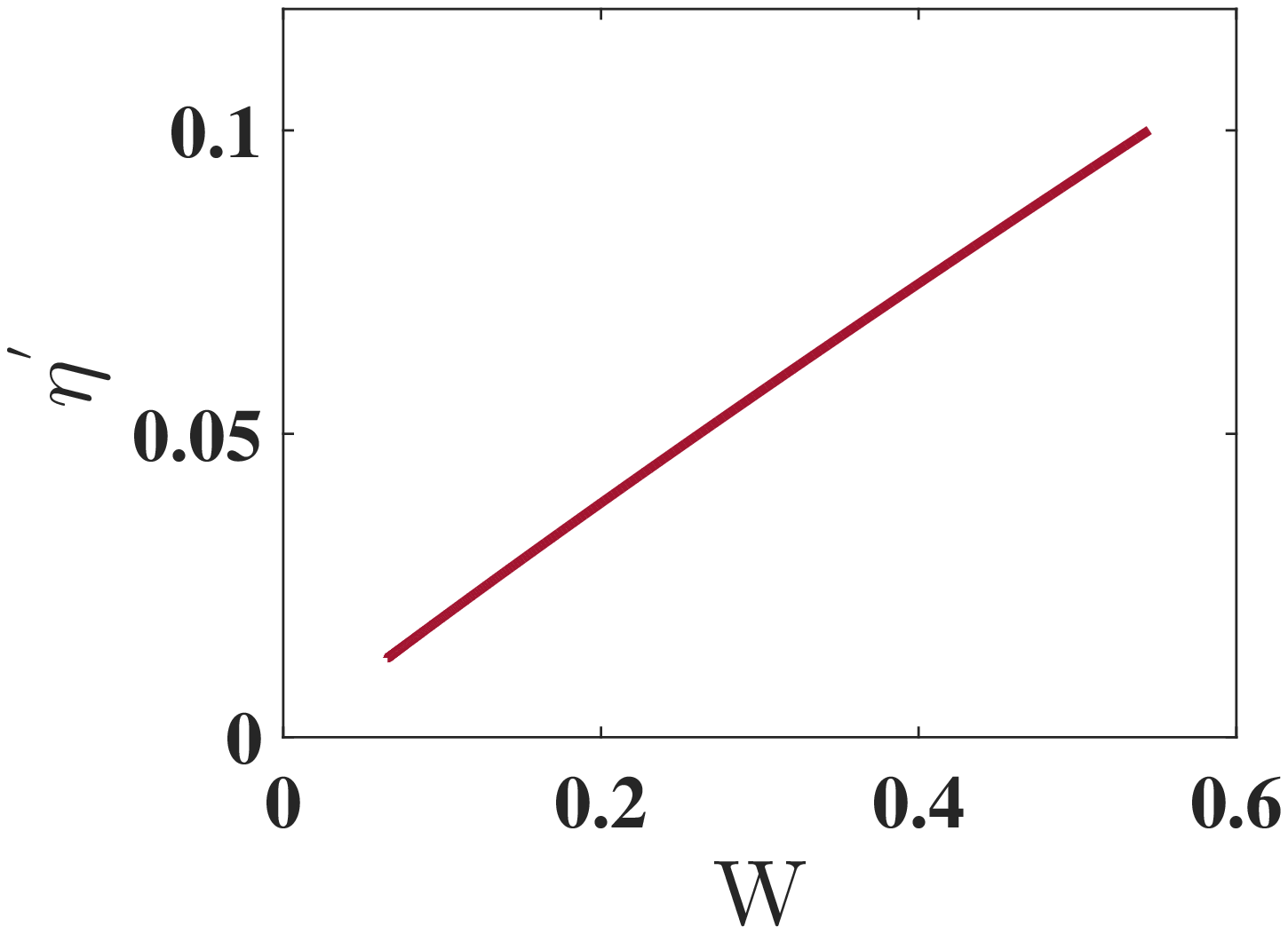}\\(a) & (b)
\end{array} $
\caption{Variation of the efficiency $\eta'$ (a) as a function of (a) coupling constant $J$ and (b) work $W$ done by the system, when the system in measured in $|E_1\rangle$ state. In (b), $J$ is changed from 0 to $2B_{L}$. The others parameters are the same as in Fig. 2. }
\end{figure}

As an example, we show in Fig. 7 how the efficiency $\eta'$ varies with $J$ and the work $W$ done by the system, when the system is measured in the state $|E_1\rangle$. Clearly, the achievable efficiency becomes less than that in Fig. 2(b), considering the effect of measurement cost.  

\section{Conclusion}\label{s:v}
We have shown how two interacting trapped ions can be employed to perform as a quantum Otto machine. These ions interact with a thermal bath at an equilibrium temperature $T_H$, while the common vibrational mode of the ions is chosen as the relevant cold bath. In order to perform the adiabatic stroke of the Otto cycle, we change the local magnetic field adiabatically. A projective measurement of the electronic states during one of the isochoric strokes leads to heat exchange with the cold bath. We find that by suitable choice of the projected state, one can effect either heat release to the cold bath or heat absorption from the cold bath, thereby leading to a heat engine or a refrigerator operation. Particularly speaking, projection onto the ground state of the system Hamiltonian results in a heat engine, while that onto the other states leads to refrigeration. The efficiency of the heat engine or the coefficient of performance of the refrigerator depends on the magnetic fields and the interaction strength between the two ions.  We assess the performance of these heat machines, by including the measurement cost, as a function of the interaction strength. We emphasize that our model is feasible with the current trapped-ion technology, as we do not need to switch off the interaction with any of the baths during the heat cycle, and still can mimic all the heat strokes of a standard Otto cycle.
%
%

\end{document}